\documentstyle[aps,preprint]{revtex}
\begin{document}

\baselineskip=20pt

\thispagestyle{empty}

\vspace{20mm}

\begin{center}
\vspace{5mm}
{\Large \bf
The $p_T$ Distribution of $J/\psi$ in QGP}

\vspace{2cm}

{\large 
Xiao-Fei Zhang$^{a}$, Han-Wen Huang$^{b}$, Xue-Qian Li$^{c,d}$,\\
and Wei-Qin Chao$^{a,c}$}\\[2mm]

{\small $^{a}$Institute of High Energy Physics, \\
Academia Sinica,
P.O.Box 918-4, 100039, P.R.China} \\
{\small $^{b}$Institute of Theoretical Physics, Academia Sinica, \\
P.O.Box 2735, Beijing 100080, China.}\\
{\small $^{c}$ CCAST (World Laboratory), Beijing, 100080, P.R. China }\\
{\small $^{d}$Department of Physics, Nankai University, Tianjin 300071,
China.}\\[20mm]
\begin{minipage}{100mm}
\centerline{\bf Abstract}
\vspace{5mm}

We consider the contributions of both color-singlet and octet $c\bar c$ bound
states for studying the suppression of $J/\psi$ production in quark gluon
plasma (QGP) and have found that the existence of the color-octet $c\bar c$
may cause the $p_T-$distribution of $J/\psi$ a decrease with $p_T$ after a
critical value $(p_T)_{max}$, which is quite different from the behavior
in Hadronic Matter (HM).

\end{minipage}
\end{center}

\vspace{20mm}
PACS number(s): 11.30.Er, 13.85.Qk, 12.15.Cc

\newpage

\noindent{I. Introduction}

It was expected that a suppression of $J/\psi$ production in
relativistic heavy ion collisions can serve as a clear signature for
the formation of a new matter phase Quark Gluon Plasma (QGP) \cite{Matsui}.
This suppression effect was observed by NA38 collaboration later\cite{NA38}.
However, successive research pointed out that such suppression could also
exist in Hadronic Matter (HM), even though by a completely different
mechanism \cite{Capp}. Therefore it is believed that an observation
of mere $J/\psi$ production rate suppression is not sufficient to confirm QGP
formation, and a more thorough study on the characteristics of
the $J/\psi$ suppression is necessary and may lead to final
confirmation of QGP state. The $p_T$
distribution of $J/\psi$  may be a good candidate, 
since the different
mechanisms which result in $J/\psi$ suppression in both
HM and QGP may provide different $p_T$ distributions of 
$J/\psi$\cite{static}\cite{ev}\cite{gpt}\cite{spt}.
 Our numerical results confirm
this conjecture (see below).

In this work, we discuss the $p_T$ distribution of $J/\psi$ in QGP and
analyze differences from that in HM.

In principle, the $J/\psi$ quarkonium is described in a Fock state
decomposition
\begin{eqnarray}
\label{Fock}
|J/\psi> &=& O(1)|c\bar c(^3S_1^{(1)})>+O(v)|c\bar c(^3P_J^{(8)}) g>
+O(v^2)|c\bar c(^1S_0^{(8)})g>+ \nonumber \\
&& O(v^2)|c\bar c(^3S_1^{(1,8)})gg>+O(v^2)|c\bar c(^3D_J^{(1,8)})gg>+...,
\end{eqnarray}
where $^{2s+1}L_J^{(1,8)}$ charaterizes the quantum state of the $c\bar c$
with color-singlet or octet respectively. This expression is valid for
the non-relativistic  QCD (NRQCD) framework and the coefficients of each
components depend on the three-velocity
$|\vec v|$ of the heavy quark, and under the limit of
$|\vec v|\rightarrow 0$, i.e. both $c$ and $\bar c$ remain at rest,
eq.(\ref{Fock}) recovers the expression for color-singlet picture of
$J/\psi$ where $O(1)\equiv 1$.

The color-octet component $(c\bar c)_8$ has been proved to play an
important role for interpreting the CDF experimental data \cite{Braaten},(see next section for more detail), 
therefore one can expect that the
color-octet $(c\bar c)_8 $ would also manifest itself in heavy ion collisions.
 Since the color-octet can interact with gluons much more strongly than
the color-singlet $(c\bar c)_1$,
it would dissolve much faster into $D$ and $\bar D$ than $(c\bar c)_1$, while traversing  through a gluon-rich environment such as QGP,
Moreover, the linear confinement potential $\sigma r$
at finite temperature  turns out to be in the form
\begin{equation}
\label{pot}
\sigma(T)r=\sigma_0({T_{dec}-T\over T_{dec}})^{\delta}\theta(T_{dec}-T)r,
\end{equation}
where $\delta$ is a parameter.
So after $T>T_{dec}$, the linear confinement disappears and the Casimir
operator $<8|{\lambda^a\over 2}{\lambda^a\over 2}|8>=+3>0$, the Coulomb-type
potential would become repulsive for an color-octet $(c\bar c)_8$. It means
that in QGP $c$ and $\bar c$ cannot be bound together in an octet, and the
original
color-octet bound state $(c\bar c)_8$ which is produced at the
early stage of the collision would dissolve.

In other words, in principle, the color-octet $(c\bar c)_8$ does exist in HM,  
but could not be built up in QGP. However, in our phenomenological
study, the main fraction of $(c\bar c)_8$ dissolves in QGP, while a
small amount remains (see next section).

In this scenario, the $(c\bar c)_8$ does not make a substantial contribution 
in QGP, as it does in $p\bar p-$collisions where the
color-octet $(c\bar c)_8$ constitutes the dominant part for large $p_T$
distribution of $J/\psi$. Thus one would expect that the large
$p_T$ contribution of $J/\psi$ would be more suppressed than the smaller
$p_T$ one in QGP, but this does not occur for HM and we will discuss it in more details later.

In next section, we describe the color-octet scenario borrowed from the
$p\bar p$ collisions and discuss a simple but popular model for $J/\psi$
suppression in QGP. In the third section we present the numerical
results, while the last section is devoted to our conclusion and discussion.\\

\noindent{II. Formulation}

(i) The color-octet $J/\psi$.

It has  already been demonstrated that the Fock decomposition 
form of $J/\psi$ can be expressed as eq.(\ref{Fock}) 
and  the coefficients of each term are functions of the
velocities.  Calculating the coefficients has been proved 
to be beyond the present ability, since this is non-perturbative
 QCD process. In principle, 
by the information gained in $p\bar p$ collisions,
the produced $c\bar c$ can be in either color-singlet or 
color-octet.
According to the newly developed factorization formalism \cite{D51},
the inclusive $J/\psi$ production amplitude from the collision of
 partons
a, b can be factorized into short- and long-distance pieces:
\begin{equation}
{\cal A}(ab\rightarrow J/\psi+X)=\sum_{(c)L,J}{\cal A}(ab\rightarrow c\bar c(^{2s+1}
L_J^{(c)})X)\times {\cal A}(c\bar c(^{2s+1}L_J^{(c)})\rightarrow J/\psi).
\end{equation}
The short-distance part for the hard production of 
$c\bar{c}$ pair can be calculated
within perturbative QCD. On the other hand, the long-distance 
component of the hadronization $c\bar c(^{2s+1}L_J^{(c)})
\rightarrow J/\psi$ is a non-perturbative process, 
which can be calculated by lattice simulation or expressed
in terms of some phenomenological parameters.

For the $p_T$ distribution, the differential cross section 
reads \cite{Chao,Chao1}
\begin{equation}
{d\sigma(p\bar p\rightarrow J/\psi+X)\over dp_T}=
{d\sigma(p\bar p\rightarrow (J/\psi)_{octet}+X)\over dp_T}+
{d\sigma(p\bar p\rightarrow (J/\psi)_{singlet}+X)\over dp_T},
\end{equation}
where the subscripts denote different color components 
of eq.(\ref{Fock}).
At the large $p_T$ limit, the main contribution comes from gluon
fragmenting into a color-octet $c\bar c$ pair.

It is important to note that in hadron collisions, 
the non-perturbative
process of evolution from the produced $c\bar c$ to $J/\psi$ 
is expressed
in terms of some phenomenological parameters which are 
independent of $p_T$,
whereas in hadron-nucleus and nucleus-nucleus collisions, 
the hadronization
of $c\bar c$ depends on $p_T$. In fact, we take the parameters 
achieved in
hadron collision to be given and consider the dependence can be 
obtained 
automatically for ion collision situation(see below).

With the understanding, we can assume that the color-singlet 
and octet $c\bar c$ are produced at early stage of
heavy ion collisions and later evolve
according to the environment. If the temperature and density 
have not reached the phase transition, the system remains 
in HM, but if the phase transition
indeed occurs, a QGP phase is formed, thus the $(c\bar c)_1$ and 
$(c\bar c)_8$ would have different evolution rules. 
It is noted that here we assume, the
$c\bar c$ are produced at the early stage of the ion 
collisions through hard N-N collisions
when the phase transition has not occurred yet, and then
the system would either remains in HM or a new phase QGP 
is formed at later
time. Thus we only need to deal with the evolution of 
$(c\bar c)$. Namely
we ignore the production of $c\bar c$ by the deconfined 
quarks or gluons in QGP.\\

(ii) A typical mechanism for suppression of $J/\psi$ production in QGP.

The $p_T$ distribution of $J/\psi$ can be described by a fundamental
picture\cite{static}\cite{ev}. Due to
the Debye screening, the potential between $c$ and $\bar c$ would 
undergo a modification from $V_{Coul}={\alpha_{eff}\over r}$ into
${\alpha_s\over r}e^{-r/r_D}$  where $\alpha_{eff}\propto \alpha_s
<1,8|\lambda^a\lambda^a|1,8>$, and when $r_D$ is sufficiently small,
$(c\bar c)_1$ can no longer form  $J/\psi$. In addition,  
as discussed above,  due to the repulsive interaction between 
$c\bar c$ in octet state in QGP,
the $(c\bar c)_8$ produced at the early stage of the collision 
can in principle, never substantialize into $J/\psi$ in  the QGP 
background if there is enough time or the size of the QGP region
 is sufficiently large. But we will see,  it is not always the case.
As for the case in HM case, $(c\bar c)_1$ and $(c\bar c)_8$ 
also have possibility to dissolve into $D\bar D$ pairs by 
scattering with the hadronic matter to reduce the $J/\psi$ 
production rate. It is noted that the mechanisms in
HM and QGP for the dissolution are different, for the 
color-singlet and octet they lead to even more apart results.

To concretely calculate the changes, let us take a reasonable and 
simplified physical picture\cite{static}\cite{ev} . 
Provided that in  A-A collisions, the initially
produced bound state $(c\bar c)_1$ has a very small radius and needs
approximately $\tau_{\psi}\sim 0.7$ fm to reach the intrinsic 
size of $J/\psi$. Thus the $(c\bar c)_1$ bound state of small 
volume does not suffer from the Debye screening in QGP. In the
 laboratory frame, due to time dilation, one has
$t_{\psi}=\tau_{\psi}\sqrt{1+(p_T/M)^2}, $
which turns out to be the formation time assuming $p_z=0$.
For the case of $p_z\neq 0$, this result applies to the transverse
distribution. Therefore if the
bound state $(c\bar c)_1$ emerges from the QGP region, within a time
interval  $t_{f}$  less than $t_{\psi}$, it can ignore 
the QGP effect  and eventually
forms $J/\psi$ meson which can be observed experimentally. 
There are two models, the static\cite{static} and the plasma 
expansion\cite{ev}. Because
the QGP phase is produced at RHIC and evolves very fast, we take the
second model to study the $p_T$ distribution of $J/\psi$. For 
$(c\bar c)_8$,
as the potental is repulsive in  QGP, $(c \bar c)_8$ suffers the 
effect of QGP almost immidiently, so the corresponding time 
$t_{\psi}^{(8)}$ being the time for $(c\bar c)_8$ starting to 
suffer from QGP effect should be much shorter than
$t_{\psi}$ due to the strong interaction with the gluons in 
environment.

Introducing $F_1$ as the ratio of the number of produced 
$(c\bar c)_1$,
which eventually forms $J/\psi$ to  that of the total $(c\bar c)_1$
produced in collisions. Similarly, we could also define a 
ratio $F_8$ for $(c\bar c)_8$. Obviously
\begin{equation}
\left\{ \begin{array}{ll}
F_1=0 & \hspace{2cm} if\;\;\;\; t_f\geq t_{\psi},\\
F_1=1 & \hspace{2cm} if\;\;\;\; t_f<t_{\psi},
\end{array} \right.
\end{equation}
where $t_f$ is the time as $(c\bar c)_1$ lingers in QGP. By the Bjorken
hydrodynamic model,
\begin{equation}
\label{full}
F_1(p_T)={\int^{R_0}_0dr\int^{\pi}_0d\phi r\rho(r)\theta(t_{\psi}-t_f(r,\phi))
\over \pi\int^{R_0}_0dr r\rho(r)},
\end{equation}
where $\rho(r)=(1-r^2/R_0^2)^b \theta(R_0-r) $ 
with  b=1 is the initial distribution density
of $(c\bar c)_1$ which is located at $\vec r$ from the center. It is
easy to find that there is a maximum transervse momentum $(p_T)_{max}$ and
one has $F_1=1$ as long as $p_T\geq (p_T)_{max}$. 
\begin{equation}
\label{max}
(p_T)_{max}=M\sqrt{({t_m\over\tau_{\psi}})^2-1},
\end{equation}
where $t_m$ is the lifetime of the QGP and M is the mass of the $J/\psi-$meson.
 It is noted
that $F_1=1$ as long as $p_T\geq (p_T)_{max}$. Actually Eq.(7) only holds 
in cases where the plasma lives for a time short compared to the average
transit time $t_1$ of a $c \bar c$ in the system.  This is because of 
the finite size effect: a $(c\bar c)_1$ with a large $p_T$ escape 
the interaction region before it is formed\cite{static}. As has been discussed 
in Ref.\cite{ev},\cite{Bl}, 
in general, for $A-A$ collisions the transverse expansion is comparatively
small, so that the condition  $t_m<t_1$ is satisfied.

   When $d=v\cdot t_{\psi}\ll R(t_{\psi})$ where $v$ is 
the velocity of $c\bar c$, one can obtain an approximate expression
\begin{equation}
F_1=\biggl({\tau_{\psi}\over\tau_m}\sqrt{1+({p\over M})^2}\biggr)^{{b+1\over a}}
\end{equation}
with approximately $a=1/2, b=1$ for equal nuclear collision.
In this expression, the surface effects are neglected compared to
eq.(\ref{full}). For not too large $p_T$ and large nuclei, the results
by the two expressions are very close\cite{ev}\cite{Bl}. Therefore,
 we will use this approximate
one to make a qualitative discussion on the problem below.\\

(iii) The effects of $(c\bar c)_8$ on the $p_T$ distribution.

As understood, at the initial stage of the A-A collisions, just in 
analog to the hadron-hadron collisions, there are both 
$(c\bar c)_1$ and $(c\bar c)_8$
produced, by the information gained from hadron collisions, their
relative fractions are $p_T-$dependent and can be expressed as
 $f_1(p_T)$ and $f_8(p_T)$ respectively (a simplified notation 
from eq.(\ref{Fock})).
They would also make different contributions to the $p_T$ 
distribution of
$J/\psi$. Thus the ratio of the number of $(c\bar c)$
which finally become $J/\psi$ to the total number of 
originally produced $(c\bar c)$ could be expressed as
\begin{equation}
F=F_1(p_T)f_1(p_T)+F_8(P_T)f_8(p_T)
\end{equation}
where $F_1$ and $F_8$ are the corresponding ratios for $(c\bar c)_1$ and
$(c\bar c)_8$ respectively as defined above.

We have argued that since $(c\bar c)_8$ is colored, it interacts 
strongly with
gluons in the environment of QGP and can dissolve much faster than $(c\bar c)_1$
which is color-blind. One can expect that the $(c\bar c)_8$ does not
contribute to $p_T$ distribution much, namely $F_8\ll F_1$ in QGP. Moreover,
if the potential picture given in eq.(\ref{Fock}) is valid, $(c\bar c)_8$
cannot survive in QGP after all, thus $F_8=0$, if the time the $(c\bar c)_8$
needs to traverse cross the QGP region is very long, there would be no
$(c\bar c)_8$ contribution to $J/\psi$. However, this has only a
statistical meaning. 
For a finite size of the region, there should be  fluctuation that
a  small fraction of $(c\bar c)_8$ emerging out. Anyhow, all models we have so far
are phenomenological and are not derived from the first principle of QCD,
therefore one can expect some small declination from that determined by the
model. So even though unable to derive $F_8$, we can phenomenologically
choose
some reasonable ratios as $F_8=F_1/2,\; F_1/10$ and $F_8=0$ for our
numerical computations.\\

\noindent{III. The numerical results}

With our simple model and phenomenological choice of $F_8$, we have
calculated the $p_T$ distribution of $J/\psi$. The numerical results are depicted in Fig.1.

Because the $(p_T)_{max}$ (see eq.(\ref{max})) depends on the formation time
of $J/\psi$ and the lifetime of the QGP, its estimation cannot be accurate
in  our present knowledge, so that we take it as a free parameter and
the only one in the whole calculation. Thus one can notice that the position
of $(p_T)_{max}$ is fixed just because we take it as a parameter and
everything is scaled with this value.  We have found that changes of
$(p_T)_{max}$ do not affect our qualitative conclusion that we predict
an increase of F with increase of $p_T$ at $p_T<(p_T)_{max}$, but a linear
decrease with further increase of $p_T$ after $p_T\geq (p_T)_{max}$. The 
choices of the ration $F_8/F_1$ also do not affect our qualitative
conclusion.

More discussions will be given in next section.\\

\noindent{IV. Conclusion and discussion}

We employ a simple model to estimate the $p_T$ distribution of $J/\psi$
which emerges from a QGP region. The data of hadron collisions indicate that
the $(c\bar c)_8$ component is not negligible and especially plays a crucial
role for the large $p_T$ distribution of produced $J/\psi$.
It motivates us to consider effects of the $(c\bar c)_8$ component
 on the
$p_T$ distribution of $J/\psi$ in relativistic heavy ion collisions,
especially when the excited region turns into the QGP phase. Indeed,
$(c\bar c)_8$ produced at the early stage of collision would have a
completely different behavior of evolution in QGP and HM. It is easy to
understand because QGP phase provides
a gluon-rich environment which strongly interacts
with the color-octet $(c\bar c)_8$ while there are no free gluons in the HM
background, thus the interaction is much weaker. Therefore the $p_T$
distribution may be taken as a signature of QGP formation.

Since the mechanism for $J/\psi$ suppression in QGP is obviously
simpler than that in HM, so we first study the $p_T$ distribution of $J/\psi$
emerging from the QGP region by calculating the F-value while only keep a
qualitative analysis of corresponding observation for HM.

Our numerical results and qualitative arguments all indicate that as
$p_T>(p_T)_{max}$ $F-$value decreases with $p_T$ and it is in contrary to
the behavior of $F$ at HM where $F$ always increases with $p_T$.
That is significant. It is
believed that in the previous $O-U$ and $S-U$ experiments, the energy
scale is not high enough to produce a QGP state, and
then the observed $F-$value
indeed keeps increasing with an increase of $p_T$. The hadron picture
gives a satisfactory description to the result \cite{Bl}\cite{Ru}, so it confirms
that a turning over of the dependence of $F$ on $p_T$ may signify a QGP
formation.

An unusual suppression of $J/\psi$ has been observed at present $Pb-Pb$
collision experiment and there have been some discussions about the
possible mechanisms \cite{Pb}. Based on our model, 
further measurement of the $p_T$-dependence of the $J/\psi$ 
suppression may provide additional information about the formation of
QGP.

As discussed above, the suppression of $J/\psi$ production also occurs at
HM, so the phenomenon is hard to be taken as a
clear signature of QGP, however, a detailed analysis on its characteristics
may do the job. The
produced $(c\bar c)_1$ and $(c\bar c)_8$ evolve according to different ways
in HM and QGP, so that the $p_T$ distribution of $J/\psi$ determined by
their contributions would be different in two matter phases. One can
optimistically hope that an analysis of the $p_T$ distribution of $J/\psi$,
especially the dependence of $F-$value on $p_T$ may inform us of existence
of QGP.

However, to make a convincing conclusion, one should do a quantitative
estimation of $F-$value at HM and QGP and this is our goal of next work
\cite{Zhang}. We believe that we cannot confirm the QGP formation even
though we observe an F-decrease with $p_T$ unless we have a thorough
study on its behavior at HM. All these not only require more theoretical
work which we will publish elsewhere, but also demand more accurate
data which will be given by  RHIC experiments.

\vskip 5mm
\noindent
{\Large{\bf Acknowledgments}}
\vskip 5mm
This work was supported in part by the National
Natural Science Foundation of China.

\vspace{2cm}

\noindent
\centerline{\bf Figure Captions}

Fig.1, the dependence of the $F-$values on $p_T$, where different curves
correspond to various ratios of $F_8/F_1$ which are phenomenologically
introduced. The $(p_T)_{max}$ is the only parameter in the scenario which
is discussed in context.

\vskip 5mm

\end{document}